\documentclass[twocolumn,english,superscriptaddress,showpacs,longbibliography]{revtex4-1}
\usepackage[T1]{fontenc}
\usepackage[latin9]{inputenc}
\usepackage{bm}
\usepackage{amsmath}
\usepackage{amssymb}
\usepackage{graphicx}
\usepackage{babel}
\begin{document}

\title{Dynamics of the Wigner Crystal of Composite Particles}

\author{Junren Shi}
\email{junrenshi@pku.edu.cn}

\selectlanguage{english}%

\affiliation{International Center for Quantum Materials, Peking University, Beijing
100871, China}

\affiliation{Collaborative Innovation Center of Quantum Matter, Beijing 100871,
China}

\author{Wencheng Ji}

\affiliation{International Center for Quantum Materials, Peking University, Beijing
100871, China}
\begin{abstract}
Conventional wisdom had long held that a composite particle behaves
just like an ordinary Newtonian particle. In this paper, we derive
the effective dynamics of a type-I Wigner crystal of composite particles
directly from its microscopic wave function. It indicates that the
composite particles are subjected to a Berry curvature in the momentum
space as well as an emergent dissipationless viscosity. Therefore,
contrary to the general belief, composite particles follow the more
general Sundaram-Niu dynamics instead of the ordinary Newtonian one.
We show that the presence of the Berry curvature is an inevitable
feature for a dynamics consistent with the dipole picture of composite
particles and Kohn's theorem. Based on the dynamics, we determine
the dispersions of magneto-phonon excitations numerically. We find
an emergent magneto-roton mode which signifies the composite-particle
nature of the Wigner crystal. It occurs at frequencies much lower
than the magnetic cyclotron frequency and has a vanishing oscillator
strength in the long wavelength limit. 
\end{abstract}

\maketitle

\section{Introduction}

In a two-dimensional electron gas (2DEG) subjected to a strong magnetic
field, electrons are forced into Landau levels with the kinetic energy
quenched. The dominating electron-electron interaction induces various
correlated ground states. The most celebrated of these states is the
fractional quantum Hall (FQH) liquid which occurs in the vicinity
of a set of magnetic filling factors of rational fractions~\cite{jainendrak.jain2007}.
Interestingly, even though the system is dominated by the electron-electron
interaction, its physics can be well described in a hidden Hilbert
space by a set of weakly interacting composite fermions (bosons) that
are bound states of an electron with an even (odd) number of quantum
vortices, as suggested by the theory of composite fermions (CFs)~\cite{jainendrak.jain2007,jain2009}.
In the theory, a FQH state is interpreted as an integer quantum Hall
state of CFs. The theory achieves great successes. For instance, the
ground state wave functions prescribed by the CF theory for the FQH
states achieve high overlaps with those determined from exact diagonalizations~\cite{jainendrak.jain2007},
and the predictions based on an intuitive picture of non-interacting
CFs are verified in various experiments~\cite{o.heinonen1997}. The
theory can even be applied to more exotic situations such as the half-filling
case which is interpreted as a Fermi liquid of CFs~\cite{kalmeyer1992,halperin1993},
and the $5/2$-filling case which is interpreted as a $p$-wave pairing
state of CFs~\cite{moore1991}. In effect, for every known state
of electrons, one could envision a counterpart for CFs. 

It is natural to envision that CFs may form a Wigner crystal (WC).
Electrons form a Wigner crystal at sufficiently low density when the
electron-electron interaction dominates over the kinetic energy~\cite{gabrielegiuliani2005}.
In the presence of a strong external magnetic field, the kinetic energy
is completely quenched and electrons should have a tendency to form
a crystalline phase. However, in 2DEG, the tendency is preempted by
the more stable FQH states when the filling factor is close to special
fractions such as $1/3$ and $2/5$. Nevertheless, a WC could be stabilized
when the filling factor deviates from these fractions. Theoretical
studies suggest that the WC of composite particles (CPWC), \emph{i.e.},
a WC consisting not of electrons but composite fermions or bosons,
could be stabilized~\cite{yi1998}. More specifically, either a type-I
CPWC~\cite{yi1998,narevich2001}, in which all composite particles
(CPs) are frozen, or a type-II CPWC~\cite{archer2013}, which lives
on top of a FQH state and only freezes CFs excessive for the filling
fraction of the FQH state, could be energetically favored over the
ordinary electron WC~\cite{yi1998,yang2001,lee2002,goerbig2004,chang2005,chang2006}.
Experimentally, there had accumulated a large number of evidences
indicating the formations of WCs in 2DEG systems, although these experiments,
either detecting the microwave resonances of disorder pinning modes~\cite{andrei1988,engel1997,li1997,li2000,chen2003,chen2004,chen2006,zhu2010,williams1991}
or measuring transport behaviors~\cite{zhang2015,liu2014,pan2002,li1991,jiang1991,williams1991,jiang1990,goldman1990},
cannot unambiguously distinguish a CPWC from its ordinary electron
counterpart.

A possible way to distinguish a CPWC from its ordinary electron counterpart
is to examine its low energy phonon excitation. The phonon excitation
of an ordinary WC had been thoroughly investigated~\cite{maki1983,c^ote1990,c^ote1991}.
It consists of a low-frequency branch and a magneto-plasmon mode that
occurs near the magnetic cyclotron frequency. Moreover, Kohn's theorem
asserts that the magnetic cyclotron mode exhausts all the spectral
weight in the long wavelength limit~\cite{kalmeyer1992,simon1998}.
For a CPWC, in analog to the ordinary WC, one would expect that its
phonon excitation also consists of two branches. However, similar
to the magneto-roton mode arisen in FQH liquids~\cite{girvin1986},
its high-frequency branch must be an emergent mode originated purely
from the electron-electron interaction, irrelevant to the cyclotron
resonance because all excitations of CPs are limited within a partially
filled Landau level. To be consistent with Kohn's theorem, the mode
must have a vanishing oscillator strength in the long wavelength limit.
These features of the emergent mode make it unique. An experimental
probe of the mode would provide an unambiguous evidence for the CP
nature of an observed WC phase.

To determine the low-energy phonon excitation of a CPWC, it is necessary
to understand the dynamics of CPs. Unfortunately, up to now, the true
nature of the CP dynamics is yet to be fully clarified. Existing theories
are based on a heuristic approach assuming that CPs follow the ordinary
dynamics characterized by an effective mass and an effective magnetic
field~\cite{rhim2015,archer2011,ettouhami2006}. The assumption is
in accordance with the conventional wisdom that a CP behaves just
like an ordinary Newtonian particle, as implied in either the Halperin-Lee-Read
theory of composite Fermi-liquids~\cite{halperin1993} or L\'{o}pez-Fradkin's
construction of the Chern-Simons field theory for FQH states~\cite{lopez1991}.
However, the validity of the assumption is questionable. An indication
of that is the violation of Kohn's theorem: the heuristic approach
would predict a new cyclotron mode corresponding to the effective
mass and magnetic field. It hints that the CP dynamics adopted by
the heuristic approach cannot be the correct one. Actually, even for
electrons in a solid, the dynamics in general has a symplectic form
(Sundaram-Niu dynamics) in which Berry curvature corrections emerge
as a result of the coupling between orbital and internal degrees of
freedom (e.g., spin-orbit coupling, SOC)~\cite{sundaram1999,xiao2010}.
It was also found that the lattice dynamics of a magnetic solid with
a strong SOC is subjected to an emergent gauge field~\cite{qin2012}
which gives rise to a dissipationless viscosity~\cite{hughes2011}.
For the case of the CPs, although the SOC is irrelevant, their orbital
motions are nevertheless entangled with internal degrees of freedom
in a strongly correlated ground state. It is reasonable to expect
that the entanglement would serve as an effective SOC and give rise
to Berry curvature corrections to the dynamics of CPs. Recently, Son
also questions the validity of the assumption by noting inconsistencies
in the conventional Halperin-Lee-Read theory of CF Fermi liquids,
and hypothesizes that a CF could be a Dirac particle~\cite{son2015,son2016}.
A Dirac particle would follow a dynamics subjected to a Berry curvature
in the momentum space with a singular distribution. 

We believe that a concrete answer to the question should be a derivation
of the CP dynamics directly from microscopic wave functions. The theory
of CFs, as detailed in Ref.~\cite{jainendrak.jain2007}, is not only
an intuitive picture for describing FQH states, but also a systematic
way for constructing ground state wave functions as well as the Hilbert
space of low-lying excitations. These informations are sufficient
for an unambiguously determination of the dynamics. Conversely, a
proposal on the nature of CFs should have an implication on how the
microscopic states would be constructed. Unfortunately, the correspondence
between the microscopic states and the dynamics is rarely explicitly
demonstrated in literatures. A rare example of the correspondence
can be found in the dipole picture of CFs~\cite{read1994}, which
is based on the microscopic Rezayi-Read wave function for a CF liquid~\cite{rezayi1994}.
However, even in this case, an explicit form of the dynamics had never
been properly formulated (see Sec.~\ref{subsec:Dipole-interpretation}). 

In this paper, we derive the effective dynamics of CPs directly from
microscopic wave functions. The derivation is based on the time-dependent
variational principle~\cite{p.kramer1981}. We focus on the type-I
CPWC, which is relatively simple without unnecessary obscuring complexities.
Based on the dynamics, we conclude that a CP, at least in the CPWC
phase, is neither an ordinary Newtonian particle nor a Dirac particle,
but a particle subjected to Berry curvature corrections, and follows
the more general Sundaram-Niu dynamics. We further show that the CP
dynamics is consistent with the dipole picture of CFs as well as Kohn's
theorem. We carry out numerical simulations to quantitatively determine
the dispersions of phonons. We find an emergent magneto-roton mode
which signifies the CP nature of a WC. The mode occurs at frequencies
much lower than the magnetic cyclotron frequency, and has a vanishing
oscillator strength in the long wavelength limit, consistent with
Kohn's theorem. The quantitative results will be useful for future
experimental probes of the emergent magneto-roton mode, which would
unambiguously distinguish a CPWC from its ordinary electron counterpart.

The remainder of the paper is organized as follows. In Sec.~II, we
derive the CP dynamics from the microscopic wave function of the CPWC
phase. In Sec.~III, we analyze CP pictures emerged from the dynamics.
In Sec. IV, we carry out the numerical simulations based on the formalism,
and present quantitative results for the dispersions of the phonon
excitations. Finally, Sec.~V contains concluding remarks.

\section{CP dynamics in a CPWC}

\subsection{CPWC wave function}

The theory of CFs prescribes an ansatz for constructing the wave functions
of the ground state and low-lying excited states of a CF system. A
CF wave function is derived from a Hartree-Fock wave function $\Psi_{HF}$,
which describes the quantum state of a collection of weakly interacting
particles in a fictitious (hidden) Hilbert space. The CF wave function
is obtained by a transformation from $\Psi_{HF}$~\cite{jainendrak.jain2007,jain2009}:
\begin{equation}
\Psi\left(\left\{ \bm{r}_{i}\right\} \right)=\hat{P}_{LLL}J\Psi_{HF}\left(\left\{ \bm{r}_{i}\right\} \right),\label{eq:PsiMap}
\end{equation}
where $\hat{P}_{LLL}$ denotes the projection to the lowest Landau
level (LLL), and 
\begin{equation}
J=\prod_{i<j}\left(z_{i}-z_{j}\right)^{m},\label{eq:Bijl-Jastrow}
\end{equation}
is the Bijl-Jastrow factor which binds an integer number of $m$ quantum
vortices to each of electrons, $z_{i}=x_{i}+iy_{i}$ with $\bm{r}_{i}\equiv(x_{i},y_{i})$
being the coordinate of an electron~\footnote{We have assumed that the direction of $q\bm{B}$ is along $+\hat{z}$
direction, where $q$ is the charge of a carrier, and $\hat{z}$ is
the normal direction of the 2DEG plane. For the opposite case, one
should define $z_{i}=x_{i}-iy_{i}$ instead.}. Equation~(\ref{eq:PsiMap}) maps a state in the conventional Landau-Fermi
paradigm to a CF state. Using different Landau-Fermi states and following
the ansatz, it is possible to construct a whole array of CF wave functions
corresponded to various states observed in 2DEG. For instance, a set
of filled Landau levels is mapped to a FQH state~\cite{jainendrak.jain2007},
a Fermi liquid to a CF Fermi liquid~\cite{kalmeyer1992,halperin1993},
a $p$-wave superconductor to the Moore-Read state~\cite{moore1991}. 

For the ground state of a type-I CPWC, $\Psi_{HF}$ is chosen to be~\cite{yi1998}:
\begin{equation}
\Psi_{HF}\left(\left\{ \bm{r}_{i}\right\} \right)=\hat{\mathcal{A}}\prod_{i}\phi_{\bm{R}_{i}^{0}}(\bm{r}_{i}),\label{eq:PsiHF}
\end{equation}
where $\phi_{\bm{R}_{i}^{0}}(\bm{r}_{i})\propto\exp[-(\bm{r}_{i}-\bm{R}_{i}^{0})^{2}/4l_{B}^{2}-i(\hat{z}\times\bm{r}_{i})\cdot\bm{R}_{i}^{0}/2l_{B}^{2}]$
is the wave function of a LLL coherent state centering at $\bm{R}_{i}^{0}$~\cite{maki1983},
$\left\{ \bm{R}_{i}^{0},\,i=1\dots N\right\} $ forms a two-dimensional
triangular lattice, $\hat{\mathcal{A}}$ denotes the (anti-) symmetrization
of the wave function, and $l_{B}\equiv\sqrt{\hbar/eB}$ is the magnetic
length for the external magnetic field $B$. We note that $\Psi_{HF}$
is actually a trial wave function for an ordinary electron WC in the
LLL~\cite{maki1983}. The mapping of Eq.~(\ref{eq:PsiMap}) transforms
it to a trial wave function for the CPWC with a variational parameter
$m$. Different from the usual CF wave functions, $m$ for a CPWC
wave function can be even (CF) or odd (composite boson). This is because
electrons in a CPWC are spatially localized and not sensitive to the
exchange symmetry. Extensive numerical simulations based on the trial
wave function had been carried out in Ref.~\cite{yi1998}. It shows
that a type-I CPWC is indeed energetically favored over the ordinary
electron WC. 

The low-lying excited states can be constructed by modifying $\Psi_{HF}$.
An apparent modification is to replace $\left\{ \bm{R}_{i}^{0}\right\} $
with $\left\{ \bm{R}_{i}\equiv\bm{R}_{i}^{0}+\bm{u}_{i}\right\} $
which introduces deviations of the particles from their equilibrium
positions. Another physically motivated modification is to introduce
a momentum for each particle. This can be achieved by replacing $\phi_{\bm{R}}(\bm{r})$
with $\phi_{\bm{R}}(\bm{r})\exp(i\bm{k}\cdot\bm{r})$, as apparent
for a localized wave packet with a momentum $\bm{p}=\hbar\bm{k}$.
We note that the similar approach is also adopted in constructing
Rezayi-Read's wave function for a CF Fermi liquid~\cite{rezayi1994}
as well as in Girvin-MacDonald-Platzman theory of magneto-roton in
FQH liquids~\cite{girvin1986}. The modifications result in a wave-function
parameterized in $\left\{ \bm{R}_{i}\right\} $ and $\left\{ \bm{k}_{i}\right\} $:
\begin{equation}
\Psi\left(\left\{ \bm{r}_{i}\right\} \right)\propto\mathcal{A}\hat{P}_{LLL}\prod_{i<j}(z_{i}-z_{j})^{m}\prod_{i}\phi_{\bm{R}_{i}}(\bm{r}_{i})e^{i\bm{k}_{i}\cdot\bm{r}_{i}},\label{eq:wavefunction}
\end{equation}
which specifies a sub-manifold in the Hilbert space. We assume that
the ground state and low-lying phononic excited states of a CPWC completely
lie in the sub-manifold.

Following the standard procedure of applying the projection to the
LLL~\cite{jainendrak.jain2007}, we obtain the explicit form of the
wave function (\ref{eq:wavefunction}):
\begin{equation}
\Psi\left(\left\{ \bm{r}_{i}\right\} \right)\propto\mathcal{A}\prod_{i<j}(z_{i}+ik_{i}l_{B}^{2}-z_{j}-ik_{j}l_{B}^{2})^{m}\prod_{i}\phi_{\bm{R}_{i}}(\bm{r}_{i}),\label{eq:wf2}
\end{equation}
where $k_{i}\equiv k_{xi}+ik_{yi}$, and we have made a substitution
$\bm{R}_{i}+\bm{k}_{i}l_{B}^{2}\times\hat{z}\rightarrow\bm{R}_{i}$,
and dropped irrelevant normalization and phase factors. We will base
our derivation of the CP dynamics on the ansatz wave function Eq.~(\ref{eq:wf2}). 

The physical meaning of the momentum $\hbar\bm{k}_{i}$ becomes apparent
in Eq.~(\ref{eq:wf2}). It shifts $z_{i}$ in the Bijl-Jastrow factor
to $z_{i}^{v}\equiv z_{i}+ik_{i}l_{B}^{2}$. One could interpret $z_{i}^{v}$
as the position of quantum vortices binding with $i$-th electron.
The momentum is actually the spatial separation of the electron and
the quantum vortices in a CP. This is exactly the dipole picture of
CFs proposed by Read~\cite{read1994}. We note that the momentum
degrees of freedom only present in systems with $m\ne0$. For an ordinary
WC with $m=0$, the momentums have no effect to the wave function
except introducing a re-parametrization to $\left\{ \bm{R}_{i}\right\} $.
Therefore, the momentums are emergent degrees of a CP system.

When adopting the ansatz wave function Eq.~(\ref{eq:wf2}), we basically
assume that the CPWC state belongs to the same paradigm as that for
FQH states. Viewed from the new CF paradigm, the modifications introduced
in Eq.~(\ref{eq:wf2}) are well motivated in physics, notwithstanding
its highly nontrivial form. The paradigm of CFs, which dictates how
the ground state and low-lying excited states are constructed, had
been extensively tested in literatures for FQH states and others~\cite{jainendrak.jain2007}.
It is reasonable to believe that the CPWC also fits in with the paradigm.
This can be tested by comparing the wave-functions generated by the
ansatz with those obtained by diagonalizing microscopic Hamiltonians.
In this paper, we will not carry out the test. Instead, we will focus
on an immediate question, \emph{i.e.}, had one adopted the paradigm
\emph{per se}, what would be the dynamics?

It can be shown that our ansatz wave-function approach is equivalent
to a CF diagonalization~\cite{jainendrak.jain2007} (see Sec.~\ref{subsec:Quantum-Correspondence-of}
and \ref{subsec:Projection-of-the}). The equivalence could serve
as a justification for our approach. Our approach is advantageous
in the sense that it provides direct knowledge of the dynamics of
CPs, whereas the CF diagonalization technique provides an efficient
machinery for systematically improving calculations but little information
about the dynamics.

\subsection{Derivation of the CP dynamics}

To determine the dynamics of CPs in a CPWC, we employ the time-dependent
variational principle of quantum mechanics. It minimizes an action
$S\equiv\int_{t_{i}}^{t_{f}}Ldt$ with the Lagrangian~\cite{p.kramer1981}:
\begin{equation}
L=\frac{i\hbar}{2}\frac{\left\langle \Psi\left|\dot{\Psi}\right.\right\rangle -\left\langle \left.\dot{\Psi}\right|\Psi\right\rangle }{\left\langle \Psi\left|\Psi\right.\right\rangle }-V_{ee},\label{eq:Lagrangian}
\end{equation}
where we assume that the wave function depends on the time through
its parameters $\left\{ \bm{R}_{i},\bm{k}_{i}\right\} $, $V_{ee}\equiv\langle\Psi|\hat{V}_{ee}|\Psi\rangle/\left\langle \Psi\left|\Psi\right.\right\rangle $
is the expectation value of the electron-electron interaction $\hat{V}_{ee}$,
and the kinetic part of the microscopic Hamiltonian of the system
is ignored since it is quenched in the LLL. A minimization of the
action will result in a semi-classical equation of motion~\cite{sundaram1999}.
Alternatively, one could interpret the action as the one determining
the path integral amplitude of a quantum evolution in the sub-manifold
of the Hilbert space~\cite{xiao2005}. The two interpretations are
corresponding to the classical and quantum version of the same dynamics,
respectively.

We proceed to determine the explicit form of the Lagrangian. The Lagrangian
can be expanded as 
\begin{equation}
L=\sum_{i}(\bm{A}_{\bm{u}_{i}}\cdot\dot{\bm{u}}_{i}+\bm{A}_{\bm{k}_{i}}\cdot\dot{\bm{k}}_{i})-V_{ee},\label{eq:L}
\end{equation}
where $\bm{A}_{\bm{u}_{i}}$ and $\bm{A}_{\bm{k}_{i}}$ are Berry
connections in the parameter space, $\bm{A}_{\bm{u}_{i}}=-\hbar\mathrm{Im}\left\langle \Psi\left|\partial\Psi/\partial\bm{u}_{i}\right.\right\rangle /\left\langle \Psi\left|\Psi\right.\right\rangle $
and $\bm{A}_{\bm{k}_{i}}=-\hbar\mathrm{Im}\left\langle \Psi\left|\partial\Psi/\partial\bm{k}_{i}\right.\right\rangle /\left\langle \Psi\left|\Psi\right.\right\rangle $,
respectively. By using Eq.~(\ref{eq:wf2}), it is straightforward
to obtain:
\begin{align}
\bm{A}_{\bm{u}_{i}} & =-\frac{\hbar}{2l_{B}^{2}}\left\langle \hat{\bm{r}}_{i}\right\rangle \times\hat{z},\\
\bm{A}_{\bm{k}_{i}} & =-m\hbar l_{B}^{2}\left\langle \sum_{j\ne i}\frac{\bm{r}_{i}-\bm{r}_{j}+\hat{z}\times(\bm{k}_{i}-\bm{k}_{j})l_{B}^{2}}{\left|\bm{r}_{i}-\bm{r}_{j}+\hat{z}\times\left(\bm{k}_{i}-\bm{k}_{j}\right)l_{B}^{2}\right|^{2}}\right\rangle ,\label{eq:Ak1}
\end{align}
where $\left\langle \dots\right\rangle \equiv\left\langle \Psi\left|\dots\right|\Psi\right\rangle /\left\langle \Psi\left|\Psi\right.\right\rangle $,
and we ignore the anti-symmetrization in the wave function Eq.~(\ref{eq:wavefunction}).
The anti-symmetrization can be re-impose when formulating the quantum
version of the dynamics. For the case of CPWCs, the effect due to
the non-distinguishability of electrons turns out to be negligible~\cite{yi1998}. 

The Berry connections could be simplified. We make use the identity:
\begin{multline}
\bm{\nabla}_{\bm{r}_{i}}\left|\Psi\right|^{2}=-\frac{\bm{r}_{i}-\bm{R}_{i}}{l_{B}^{2}}\left|\Psi\right|^{2}\\
+2m\sum_{j\ne i}\frac{\bm{r}_{i}-\bm{r}_{j}+\hat{z}\times(\bm{k}_{i}-\bm{k}_{j})l_{B}^{2}}{\left|\bm{r}_{i}-\bm{r}_{j}+\hat{z}\times\left(\bm{k}_{i}-\bm{k}_{j}\right)l_{B}^{2}\right|^{2}}\left|\Psi\right|^{2}.\label{eq:DPsi}
\end{multline}
Substituting Eq.~(\ref{eq:DPsi}) into (\ref{eq:Ak1}), we obtain,
\begin{equation}
\bm{A}_{\bm{k}_{i}}=-\frac{\hbar}{2}\left(\left\langle \hat{\bm{\xi}_{i}}\right\rangle -\bm{u}_{i}\right),
\end{equation}
where $\hat{\bm{\xi}_{i}}\equiv\hat{\bm{r}}_{i}-\bm{R}_{i}^{0}$.
The Berry connections can then be expressed as:
\begin{eqnarray}
\bm{A}_{\bm{u}_{i}} & = & -\frac{\hbar}{2l_{B}^{2}}\bm{x}_{i}\times\hat{z}+\frac{\hbar\bm{k}_{i}}{2},\label{eq:Au}\\
\bm{A}_{\bm{k}_{i}} & = & -\frac{\hbar}{2}\left(\bm{x}_{i}-\bm{u}_{i}+\bm{k}_{i}\times\hat{z}l_{B}^{2}\right),\label{eq:Ak}
\end{eqnarray}
where $\bm{x}_{i}\equiv\left\langle \hat{\bm{\xi}_{i}}\right\rangle -\bm{k}_{i}\times\hat{z}l_{B}^{2}$,
$\hat{z}$ is the unit normal vector of the 2DEG plane . We note that
$\bm{x}_{i}$ is the average position (relative to $\bm{R}_{i}^{0}$)
of the quantum vortices binding with $i$-th electron, which is displaced
from the electron position $\langle\hat{\bm{\xi}}_{i}\rangle$ by
a vector $-\bm{k}_{i}\times\hat{z}l_{B}^{2}$, according to the wave
function Eq.~(\ref{eq:wf2}). 

We adopt $\{\bm{x}_{i},\bm{p}_{i}\equiv\hbar\bm{k}_{i}\}$ as the
set of dynamic variables, and interpret $\bm{x}_{i}$ and $\bm{p}_{i}$
as the position and momentum of a CP, respectively. To express the
Lagrangian in $\{\bm{x}_{i},\bm{p}_{i}\}$, it is necessary to relate
the dynamic variables with the original set of parameters. We assume
that both $\bm{x}_{i}$ and $\bm{p}_{i}$ are small in a CPWC, and
expand the Lagrangian to the second order of the dynamic variables.
For the purpose, we expand $\bm{x}_{i}$ to the linear order of the
original parameters:
\begin{equation}
x_{i\alpha}\approx\sum_{j\beta}A_{i\alpha,j\beta}u_{j\beta}+B_{i\alpha,j\beta}k_{j\beta},\label{eq:x}
\end{equation}
and,
\begin{align}
A_{i\alpha,j\beta} & \equiv\left.\frac{\partial\left\langle \hat{x}_{i\alpha}\right\rangle }{\partial u_{j\beta}}\right|_{0}=\frac{1}{l_{B}^{2}}\left\langle \hat{\xi}_{i\alpha}\hat{\xi}_{j\beta}\right\rangle _{0},\label{eq:A1}\\
B_{i\alpha,j\beta} & \equiv\left.\frac{\partial\left\langle \hat{x}_{i\alpha}\right\rangle }{\partial k_{j\beta}}\right|_{0}=-l_{B}^{2}\epsilon_{\alpha\beta}\delta_{ij}\nonumber \\
 & +\left\langle \hat{\xi}_{i\alpha}\left(2ml_{B}^{2}\sum_{l\ne j,\gamma}\epsilon_{\beta\gamma}\frac{r_{j\gamma}-r_{l\gamma}}{\left|\bm{r}_{j}-\bm{r}_{l}\right|^{2}}\right)\right\rangle _{0},
\end{align}
where $\alpha\,(\beta)=x,y$ indexes the component of the coordinate,
$\left\langle \dots\right\rangle _{0}$ denotes the expectation value
in the ground state $\Psi_{0}\equiv\left.\Psi\right|_{\bm{u}_{i},\bm{k}_{i}\rightarrow0}$,
and $\epsilon_{\alpha\beta}$ is the two-dimensional Levi-Civita symbol.
Making use the identity Eq.~(\ref{eq:DPsi}), we obtain:
\begin{equation}
B_{i\alpha,j\beta}=-l_{B}^{2}A_{i\alpha,j\gamma}\epsilon_{\gamma\beta}.\label{eq:B}
\end{equation}
Substituting (\ref{eq:A1}) and (\ref{eq:B}) into (\ref{eq:x}),
we obtain:

\begin{align}
x_{i\alpha} & =\sum_{j\beta}A_{i\alpha,j\beta}\left(\bm{u}_{j}-\bm{k}_{j}\times\hat{z}l^{2}\right)_{\beta},\label{eq:x-u}
\end{align}

Similarly, $V_{ee}$ is expanded to the second order of the dynamic
variables:
\begin{multline}
V_{ee}\approx\frac{1}{2}\sum_{i\alpha,j\beta}D_{i\alpha,j\beta}^{\bm{xx}}x_{i\alpha}x_{j\beta}+2D_{i\alpha,j\beta}^{\bm{px}}p_{i\alpha}x_{j\beta}\\
+D_{i\alpha,j\beta}^{\bm{pp}}p_{i\alpha}p_{j\beta}.\label{eq:Vee}
\end{multline}
The coefficients can be related to correlation functions (see Appendix
A):

\begin{align}
D_{i\alpha,j\beta}^{\bm{x}\bm{x}} & =\frac{1}{l_{B}^{4}}\sum_{\gamma\delta}\left\langle \left(\hat{V}_{ee}-\bar{V}_{ee}\right)\hat{\xi}_{l\gamma}\hat{\xi}_{m\delta}\right\rangle _{0}\nonumber \\
 & \times\left[A^{-1}\right]_{i\alpha,l\gamma}\left[A^{-1}\right]_{m\delta,j\beta},\label{eq:Dxx}\\
D_{i\alpha,j\beta}^{\bm{p}\bm{x}} & =-\frac{1}{\hbar}\sum_{\gamma\delta}\epsilon_{\alpha\gamma}\left\langle \frac{\partial\hat{V}_{ee}}{\partial r_{i\gamma}}\hat{\xi}_{l\delta}\right\rangle _{0}\left[A^{-1}\right]_{l\delta,j\beta},\label{eq:Dpx}\\
D_{i\alpha,j\beta}^{\bm{p}\bm{p}} & =\frac{l_{B}^{4}}{\hbar^{2}}\sum_{\gamma\delta}\epsilon_{\alpha\gamma}\epsilon_{\beta\delta}\left\langle \frac{\partial^{2}\hat{V}_{ee}}{\partial r_{i\gamma}\partial r_{j\delta}}\right\rangle _{0},\label{eq:Dpp}
\end{align}
where $\left[A^{-1}\right]$ denotes the inverse of a matrix with
elements $[A]_{i\alpha,j\beta}=A_{i\alpha,j\beta}$, and $\bar{V}_{ee}\equiv\langle\hat{V}_{ee}\rangle_{0}$.

Substituting Eqs.~(\ref{eq:Au}, \ref{eq:Ak}, \ref{eq:x-u}, \ref{eq:Vee})
into Eq.~(\ref{eq:L}), we can determine the explicit form of the
Lagrangian. Because of the translational symmetry, it is convenient
to express the Lagrangian in the Fourier transformed dynamic variables
$\bm{x}(\bm{q})\equiv1/\sqrt{N}\sum_{i}\bm{x}_{i}\exp\left(-i\bm{q}\cdot\bm{R}_{i}^{0}\right)$
and $\bm{p}(\bm{q})\equiv1/\sqrt{N}\sum_{i}\bm{p}_{i}\exp\left(-i\bm{q}\cdot\bm{R}_{i}^{0}\right)$,
where $\bm{q}$ is a wave vector defined in the Brillouin zone for
a triangular lattice. The Lagrangian can be decomposed into $L=\sum_{\bm{q}}L_{\bm{q}}$
with: 
\begin{multline}
L_{\bm{q}}=\frac{eB_{e}(\bm{q})}{2}\left(\hat{z}\times\bm{x}^{\ast}(\bm{q})\right)\cdot\dot{\bm{x}}(\bm{q})+\frac{1}{2eB}\left(\hat{z}\times\bm{p}^{\ast}(\bm{q})\right)\cdot\dot{\bm{p}}(\bm{q})\\
+\bm{p}^{\ast}(\bm{q})\cdot\dot{\bm{x}}(\bm{q})-\frac{1}{2}\left[\begin{array}{c}
\bm{x}(\bm{q})\\
\bm{p}(\bm{q})
\end{array}\right]^{\dagger}\mathcal{D}(\bm{q})\left[\begin{array}{c}
\bm{x}(\bm{q})\\
\bm{p}(\bm{q})
\end{array}\right],\label{eq:Lq}
\end{multline}
where $B_{e}(\bm{q})$ is determined by,
\begin{equation}
B_{e}(\bm{q})=\frac{B}{2}\mathrm{Tr}\mathcal{A}^{-1}(\bm{q}),\label{eq:Be}
\end{equation}
with $\mathcal{A}^{-1}(\bm{q})$ being the inverse of a $2\times2$
matrix with elements $\mathcal{A}_{\alpha\beta}(\bm{q})=\sum_{\bm{R}_{i}^{0}}A_{i\alpha,0\beta}\exp(-i\bm{q}\cdot\bm{R}_{i}^{0})$,
and 
\begin{equation}
\mathcal{D}(\bm{q})=\left[\begin{array}{cc}
\mathcal{D}^{\bm{xx}}(\bm{q}) & \mathcal{D}^{\bm{px}}(\bm{q})\\
\mathcal{D}^{\bm{px}}(\bm{q}) & \mathcal{D}^{\bm{pp}}(\bm{q})
\end{array}\right],
\end{equation}
with $\mathcal{D}^{\bm{xx}}(\bm{q})$, $\mathcal{D}^{\bm{px}}(\bm{q})$,
and $\mathcal{D}^{\bm{pp}}(\bm{q})$ being the Fourier transforms
of $D^{\bm{xx}}$, $D^{\bm{px}}$, and $D^{\bm{pp}}$, respectively. 

The equation of motion of a type-I CPWC is:
\begin{equation}
\left[\begin{array}{cc}
eB_{e}(\bm{q})\hat{z}\times & I\\
-I & \frac{1}{eB}\hat{z}\times
\end{array}\right]\left[\begin{array}{c}
\dot{\bm{x}}(\bm{q})\\
\dot{\bm{p}}(\bm{q})
\end{array}\right]=-\mathcal{D}(\bm{q})\left[\begin{array}{c}
\bm{x}(\bm{q})\\
\bm{p}(\bm{q})
\end{array}\right],\label{eq:EQM}
\end{equation}
which is the main result of this paper. Interpretations of the dynamics
and its implications to the nature of CPs will be discussed in Sec.~\ref{sec:Interpretations-of-the}.

\subsection{Quantization of the effective dynamics\label{subsec:Quantum-Correspondence-of}}

The dynamics Eq.~(\ref{eq:EQM}) could be quantized. The resulting
quantum dynamics describes the quantum evolution of the system in
the sub-manifold of the Hilbert space specified by the wave function
Eq.~(\ref{eq:wf2}). A general scheme of the quantization had been
discussed in Ref.~\cite{xiao2005}. Basically, the non-canonical
kinematic matrix in the left hand side of Eq.~(\ref{eq:EQM}) gives
rise to non-commutativity between the dynamic variables:
\begin{equation}
\left[\begin{array}{cc}
[\bm{x}^{\dagger}(\bm{q}),\bm{x}(\bm{q})] & [\bm{x}^{\dagger}(\bm{q}),\bm{p}(\bm{q})]\\{}
[\bm{p}^{\dagger}(\bm{q}),\bm{x}(\bm{q})] & [\bm{p}^{\dagger}(\bm{q}),\bm{p}(\bm{q})]
\end{array}\right]=i\hbar\left[\begin{array}{cc}
eB_{e}(\bm{q})\hat{\epsilon} & -I\\
I & \frac{1}{eB}\hat{\epsilon}
\end{array}\right]^{-1},\label{eq:commutation}
\end{equation}
where $\hat{\epsilon}$ is the $2\times2$ anti-symmetric matrix with
$[\hat{\epsilon}]_{\alpha\beta}=\epsilon_{\alpha\beta}$. The system
is governed by an effective hamiltonian $\hat{H}_{eff}=V_{ee}$ by
upgrading the dynamic variables to quantum operators. 

The system can be transformed to a phonon representation by a procedure
described in Ref.~\cite{qin2012}. We solve the generalized eigenvalue
equation:
\begin{equation}
i\omega_{\bm{q}}\left[\begin{array}{cc}
-eB_{e}(\bm{q})\hat{\epsilon} & I\\
-I & -\frac{1}{eB}\hat{\epsilon}
\end{array}\right]\psi_{\bm{q}}=\mathcal{D}(\bm{q})\psi_{\bm{q}}.\label{eq:gep}
\end{equation}
The equation gives rise to two positive frequency solutions and two
negative frequency solutions, with eigenvectors related by complex
conjugations~\cite{qin2012}. The eigenvectors are normalized by
$\bar{\psi}_{\bm{q}}\psi_{\bm{q}}=\pm1,$ where $\pm$ is for the
positive and the negative frequency solution, respectively, and
\begin{equation}
\bar{\psi}_{\bm{q}}\equiv-i\psi_{\bm{q}}^{\dagger}\left[\begin{array}{cc}
eB_{e}(\bm{q})\hat{\epsilon} & -I\\
I & \frac{1}{eB}\hat{\epsilon}
\end{array}\right].
\end{equation}

The dynamic variables can then be expressed in phonon creation and
annihilation operators:
\begin{equation}
\left[\begin{array}{c}
\bm{x}(\bm{q})\\
\bm{p}(\bm{q})
\end{array}\right]=\sum_{i\in+}\psi_{\bm{q}}^{(i)}a_{\bm{q}i}+\psi_{-\bm{q}}^{(i)\ast}a_{-\bm{q}i}^{\dagger},\label{eq:expansion}
\end{equation}
where the summation is over the two positive frequency solutions,
and $a_{\bm{q}i}$ and $a_{\bm{q}i}^{\dagger}$ are bosonic creation
and annihilation operators, respectively. One can verify that the
dynamic variables, expressed as Eq.~(\ref{eq:expansion}), do recover
the commutation relation Eq.~(\ref{eq:commutation}).

With the phonon representation, we define a coherent state as the
eigenstate of the annihilation operator:
\begin{equation}
a_{\bm{q}i}\left|\phi\right\rangle =\phi_{\bm{q}i}\left|\phi\right\rangle .
\end{equation}
In the the real space, the coherent state is interpreted as, 
\begin{equation}
\left\langle \bm{r}\left|\phi\right.\right\rangle =\frac{\Psi(\bm{r};\phi)}{\left\langle \Psi_{0}\left|\Psi\right.\right\rangle },\label{eq:coherent state}
\end{equation}
where $\Psi(\bm{r};\phi)$ is the wave function Eq.~(\ref{eq:wf2})
with the parameters substituted with values corresponding to:
\begin{equation}
\left[\begin{array}{c}
\bm{x}^{(+)}(\bm{q})\\
\bm{p}^{(+)}(\bm{q})
\end{array}\right]=\sum_{i\in+}\psi_{\bm{q}}^{(i)}\phi_{\bm{q}i},\label{eq:positive component}
\end{equation}
where the superscript $(+)$ indicates that the dynamic variables
contain positive-frequency components only~\cite{glauber1963}, and
the denominator is introduced to eliminate the time-dependent factor
of the ground-state component in the wave-function~\cite{p.kramer1981}. 

For a given phonon state, the corresponding physical wave function
can be determined by~\cite{glauber1963}:
\begin{equation}
\left\langle \bm{r}\left|\varphi\right.\right\rangle =\int\frac{\mathrm{d}\phi\mathrm{d}\phi^{\ast}}{2\pi i}e^{-\left|\phi\right|^{2}}\frac{\Psi(\bm{r};\phi)}{\left\langle \Psi_{0}\left|\Psi\right.\right\rangle }\varphi(\phi^{\ast}).
\end{equation}
For the excited state with $n$ phonons of the mode $(\bm{q},i)$,
$\varphi(\phi^{\ast})\propto\phi_{\bm{q}i}^{\ast n}$, the corresponding
physical wave function is:
\begin{equation}
\Psi_{n}(\bm{r})\propto\left.\frac{\partial^{n}}{\partial\phi_{\bm{q}i}^{n}}\frac{\Psi(\bm{r};\phi)}{\left\langle \Psi_{0}\left|\Psi\right.\right\rangle }\right|_{\phi\rightarrow0}.\,\label{eq:Psin}
\end{equation}

From Eq.~(\ref{eq:Psin}), we conclude that a one-phonon state must
be a superposition of:
\begin{equation}
\left.\frac{\partial}{\partial\bar{u}_{i}}\frac{\Psi(\bm{r})}{\left\langle \Psi_{0}\left|\Psi\right.\right\rangle }\right|_{\bar{u},k\rightarrow0},\,\,\,\left.\frac{\partial}{\partial k_{i}}\frac{\Psi(\bm{r})}{\left\langle \Psi_{0}\left|\Psi\right.\right\rangle }\right|_{\bar{u},k\rightarrow0}.
\end{equation}
They are corresponding to a set of many-body wave functions:
\begin{equation}
\hat{P}_{LLL}(z_{i}-Z_{i}^{0})\Psi_{0},\,\,\,\hat{P}_{LLL}(\bar{z}_{i}-\bar{Z}_{i}^{0})\Psi_{0}.\label{eq:basis}
\end{equation}

One can construct a quantum solution of the phonon excitation problem
by directly diagonalizing the microscopic hamiltonian in the truncated
Hilbert space span by the bases Eq.~(\ref{eq:basis}). The resulting
eigenvalue equation, after an appropriate basis transformation, is
nothing but the eigenvalue equation Eq.~(\ref{eq:gep}), with a modified
dynamic matrix. The modification to the dynamic matrix will derived
in the next subsection.

\subsection{Projected dynamic matrix\label{subsec:Projection-of-the}}

Before proceeding, we note a subtlety concerning the quantum correspondence
of the dynamics. In the derivation of the dynamics, we treat $\bm{x}_{i}$
and $\bm{p}_{i}$ as classical variables. However, when constructing
the quantum coherent states, we use only the positive-frequency components
of the dynamic variables, as shown in Eq.~(\ref{eq:positive component}).
The latter is necessary because the wave function $\left\langle \bm{r}\left|\phi\right.\right\rangle $
defined in Eq.~(\ref{eq:coherent state}) should be a superposition
of the ground state and excited states: $\left\langle \bm{r}\left|\phi\right.\right\rangle \sim\Psi_{0}+\sum_{i}\exp(-i\Delta E_{i}t)\Psi_{i}$
with $\Delta E_{i}>0$, \emph{i.e.}, it only contains positive frequency
components in its time dependence. By assuming that the wave function
is a function of the positive frequency components of the dynamic
variables, we are able to obtain a wave function consistent with the
general requirement~\cite{glauber1963}.

The consideration will introduce a modification to the harmonic expansion
of $V_{ee}$. This is because the expansion Eq.~(\ref{eq:Vee}),
which treats the dynamic variable as classical variables, includes
terms that couple two positive (negative)-frequency components of
dynamic variables. These terms induce spurious couplings between the
positive and negative frequency components, and should be dropped.
On the other hand, one can show that kinematic part of the dynamics
is not affected by the spurious coupling.

To determine the modification, we note that our wave function Eq.~(\ref{eq:wf2})
depends only on the complex variables $\bar{u}_{i}\equiv u_{xi}-iu_{yi}$
and $k_{i}\equiv k_{xi}+ik_{yi}$ . Thus, $\bar{u}_{i}$ and $k_{i}$
can be chosen to be positive-frequency functions of the time, and
a proper harmonic expansion of $V_{ee}$ should only include terms
coupling $\{\bar{u}_{i},k_{i}\}$ with their complex conjugates. To
this end, we expand $V_{ee}$ in terms of $\{\bar{\bm{u}}(\bm{q}),\bm{k}(\bm{q})\}$:
\begin{equation}
V_{ee}\approx\frac{1}{2}\sum_{\bm{q}}\left[\begin{array}{c}
\bar{\bm{u}}(\bm{q})\\
\bm{k}(\bm{q})
\end{array}\right]^{\dagger}\tilde{\mathcal{D}}(\bm{q})\left[\begin{array}{c}
\bar{\bm{u}}(\bm{q})\\
\bm{k}(\bm{q})
\end{array}\right],
\end{equation}
where $\tilde{\mathcal{D}}(\bm{q})$ is the dynamic matrix with respect
to $\{\bm{u}(\bm{q}),\bm{k}(\bm{q})\}$. To get rid of the spurious
coupling, we introduce a projected dynamic matrix:
\begin{equation}
\tilde{\mathcal{D}}^{P}(\bm{q})=\tilde{P}_{+}\tilde{\mathcal{D}}(\bm{q})\tilde{P}_{+}+\tilde{P}_{-}^{\dagger}\tilde{\mathcal{D}}(\bm{q})\tilde{P}_{-}.
\end{equation}
with
\begin{equation}
\tilde{P}_{\pm}=\left[\begin{array}{cc}
\frac{1}{2}\left(1\mp\sigma_{2}\right) & 0\\
0 & \frac{1}{2}\left(1\pm\sigma_{2}\right)
\end{array}\right],
\end{equation}
where $\sigma_{2}$ is the second Pauli matrix. 

Similarly, we can obtain a projected dynamic matrix with respect to
$\{\bm{x}(\bm{q}),\bm{p}(\bm{q})\}$ by a projection:

\begin{align}
\mathcal{D}^{P}(\bm{q}) & =P_{+}^{\dagger}(\bm{q})\mathcal{D}(\bm{q})P_{+}(\bm{q})+P_{-}^{\dagger}(\bm{q})\mathcal{D}(\bm{q})P_{-}(\bm{q}),\label{eq:DP}
\end{align}
with $P_{\pm}=U(\bm{q})\tilde{P}_{\pm}U^{-1}(\bm{q})$, where $U(\bm{q})$
is the transformation matrix relating $\{\bar{\bm{u}}(\bm{q}),\bm{k}(\bm{q})\}$
with $\{\bm{x}(\bm{q}),\bm{p}(\bm{q})\}$: $[\bm{x}(\bm{q}),\bm{p}(\bm{q})]^{T}=U(\bm{q})[\bm{u}(\bm{q}),\bm{k}(\bm{q})]^{T}$.
We have: 

\begin{align}
P_{\pm}(\bm{q}) & =\left[\begin{array}{cc}
\frac{1}{2}\left(1\mp\mathcal{A}(\bm{q})\sigma_{2}\mathcal{A}^{-1}(\bm{q})\right) & \mp i\mathcal{A}(\bm{q})\\
0 & \frac{1}{2}\left(1\pm\sigma_{2}\right)
\end{array}\right].
\end{align}

By substituting the dynamic matrix $\mathcal{D}(\bm{q})$ in Eq.~(\ref{eq:EQM})
and (\ref{eq:gep}) with $\mathcal{D}^{P}(\bm{q})$, one can show
that the eigenvalue equation becomes identical to that obtained from
the CF diagonalization with the bases Eq.~(\ref{eq:basis}). 

\section{Interpretations of the CP dynamics\label{sec:Interpretations-of-the}}

\subsection{Sundaram-Niu dynamics of CPs}

The CP dynamics, as shown in Eq.~(\ref{eq:EQM}), is distinctly different
from the one adopted in the heuristic approach, in which a CP is assumed
to be an ordinary Newtonian particle characterized by an effective
mass and a mean-field effective magnetic field~\cite{ettouhami2006,archer2011,rhim2015}.
Our CP dynamics fits in with the form of the more general Sundaram-Niu
dynamics with Berry curvature corrections. An analysis of these corrections
would provide an insight to the nature of CPs, as we will discuss
in the following.

Firstly, CPs are subjected to an emergent gauge field $\Delta B_{e}(\bm{q})\equiv B-B_{e}(\bm{q})$.
The emergent gauge field gives rise to a dissipationless viscosity,
which is a transverse inter-particle force proportional to the relative
velocity between two particles~\cite{qin2012,hughes2011}:
\begin{equation}
\bm{F}_{ij}^{(DV)}=e\Delta\mathcal{B}_{e}(\bm{R}_{i}^{0}-\bm{R}_{j}^{0})\hat{z}\times\left(\dot{\bm{x}}_{j}-\dot{\bm{x}}_{i}\right),\label{eq:DV}
\end{equation}
where $\Delta\mathcal{B}_{e}(\bm{R}^{0})\equiv\int_{BZ}d^{2}q/(2\pi)^{2}\Delta B_{e}(\bm{q})\exp(i\bm{q}\cdot\bm{R}^{0})$.
In ordinary phonon systems, the dissipationless viscosity could arise
in the presence of a strong SOC and magnetization. In CPWC, however,
it is induced by the quantum vortices attached in CPs, similar to
the Chern-Simons field emerged in FQH liquids~\cite{lopez1991,zhang1992}.
For the latter, one usually adopts a mean-field approximation that
gives rise to an effective magnetic field experienced by CPs. For
the CPWC, the mean-field approximation is equivalent to keeping only
the diagonal component of the emergent gauge field 
\begin{equation}
\Delta B\equiv\Delta\mathcal{B}_{e}(\bm{0})=-\sum_{\bm{R}_{i}^{0}\ne0}\Delta\mathcal{B}_{e}(\bm{R}_{i}^{0}),\label{eq:DeltaB}
\end{equation}
and assuming that CPs experience an effective magnetic field $B_{eff}=B-\Delta B$.
However, the mean field approximation may not be appropriate for a
CPWC since it breaks the translational symmetry.

Secondly, CPs are subjected to a Berry curvature in the momentum space
with $\Omega_{z}=1/eB$. This is a new feature of the dynamics, not
presented in the conventional theory of CFs~\cite{halperin1993,simon1998}.
The Berry curvature gives rise to an anomalous velocity, which is
well known for electron dynamics in magnetic solids with a SOC, and
is linked to the (quantum) anomalous Hall effect~\cite{jungwirth2002,haldane1988}.
Here, the Berry curvature is not induced by the SOC, but inherited
from the Landau level hosting the particles. Indeed, a Landau level,
when casted to a magnetic Bloch band, does have a uniformly distributed
Berry curvature in the momentum space with $\Omega_{z}^{(LL)}=-1/eB$~\cite{zhang2016}.
One can show that the difference in the signs of $\Omega_{z}$ and
$\Omega_{z}^{(LL)}$ is due to our assignment of the CP position to
its constituent quantum vortices (see Sec.~\ref{subsec:Definition-of-the}
and \cite{shi2017}). The presence of a Berry curvature in the momentum
space clearly indicates that a CP is neither an ordinary Newtonian
particle nor a Dirac particle. Interestingly, had the Berry curvature
survived in a half-filled CF Fermi liquid, it would give rise to a
$\pi$ Berry phase, the same as that predicted by the Dirac theory~\cite{son2015}.

Based on these discussions, we conclude that a CP is a particle following
the Sundaram-Niu dynamics.

\subsection{Dipole interpretation\label{subsec:Dipole-interpretation}}

The interpretation is not necessarily unique. It depends on the choice
of dynamic variables and physical meanings one assigns to them. Had
we interpreted $\hat{z}\times\bm{p}_{i}/eB$ as the displacement from
the electron to the quantum vortices in a CP, as indicated by the
wave function Eq.~(\ref{eq:wf2}), we would obtain the dipole interpretation
of the dynamics~\cite{read1994,simon1998}.

In the dipole interpretation, a CP is regarded as a dipole consisting
of an electron and a bundle of $m$ quantum vortices~\cite{read1994,simon1998}.
The picture and its relation to the usual position-momentum interpretation
had been discussed in Ref.~\cite{read1994}. For the interpretation,
we adopt another set of dynamic variables:
\begin{align}
\bm{x}_{i}^{e} & =\left\langle \hat{\bm{r}}_{i}\right\rangle -\bm{R}_{i}^{0}=\bm{x}_{i}-\frac{1}{eB}\hat{z}\times\bm{p}_{i},\label{eq:xie}\\
\bm{x}_{i}^{\phi} & \equiv\bm{x}_{i},
\end{align}
which are positions of the electron and the bundle of the quantum
vortices, respectively. Note that the position of a composite particle
is assigned to the position of the quantum vortices in Eq.~(\ref{eq:EQM}).

The equation of motion with respect to the new dynamic variables is:
\begin{equation}
\left[\begin{array}{cc}
e\Delta B_{e}(\bm{q})\hat{z}\times & 0\\
0 & -eB\hat{z}\times
\end{array}\right]\left[\begin{array}{c}
\dot{\bm{x}}_{\phi}(\bm{q})\\
\dot{\bm{x}}_{e}(\bm{q})
\end{array}\right]=\mathcal{D}^{\prime}(\bm{q})\left[\begin{array}{c}
\bm{x}_{\phi}(\bm{q})\\
\bm{x}_{e}(\bm{q})
\end{array}\right],\label{eq:EQM1}
\end{equation}
where $\mathcal{D}^{\prime}(\bm{q})$ is the corresponding dynamic
matrix, which can be related to $\mathcal{D}^{P}(\bm{q})$ by a transformation.

It is notable from Eq.~(\ref{eq:EQM1}) that the electron in a CP
is only coupled to the external magnetic field, while the quantum
vortices are only coupled to the emergent gauge field~\cite{potter2016}.
Although not explicitly specified in the original proposal~\cite{read1994},
the simple form of the coupling could have been expected from the
microscopic wave function Eq.~(\ref{eq:wf2}), in which the correlations
introduced in the Bijl-Jastrow factor are between coordinates of quantum
vortices.

It also becomes apparent that our dynamics is consistent with Kohn's
theorem. In the long wavelength limit $\bm{q}\rightarrow0$, both
$\mathcal{D}^{\prime}(\bm{q})$ and $\Delta B_{e}(\bm{q})$ vanish
because of the translational symmetry. As a result, the degrees of
freedom associating with $\bm{x}_{\phi}$ become degenerate. The system
will only have a trivial zero frequency mode, and no emergent mode
will be present. The behavior is exactly what would be expected from
Kohn's theorem, because the cyclotron mode, which is the only allowed
resonance at $\bm{q}=0$ according to the theorem, is an inter-Landau
level excitation, and will not appear in our dynamics which has assumed
that all excitations are within a Landau level.

From these observations, it becomes apparent that the presence of
the Berry curvature in the momentum space would be an inevitable feature
of the CP dynamics if we wanted to obtain the particular form of the
dipole picture or maintain the consistency with Kohn's theorem. Had
we assumed a vanishing Berry curvature in Eq.~(\ref{eq:EQM}), Eq.~(\ref{eq:EQM1})
would have a different form of the coupling to gauge fields, and its
right hand side would not become degenerate to be consistent with
Kohn's theorem. The presence of the Berry curvature is actually the
most important difference between the conventional Chern-Simons theory
of CPs~\cite{kalmeyer1992,halperin1993}, which is constructed from
particles residing in a free parabolic band~\cite{lopez1991,zhang1992},
and a theory directly derived from a microscopic wave function defined
in a Landau level.

\subsection{Definition of the CP position\label{subsec:Definition-of-the}}

In the position-momentum interpretation of the dynamics, there is
arbitrariness in defining the position of a CP. In Eq.~(\ref{eq:EQM}),
the position of a CP is interpreted as the position of its constituent
quantum vortices. It seems to be equally plausible to interpret the
CP position as the electron position (or, e.g., the average position
of the electron and the quantum vortices). The issue is: will a different
choice affect our interpretation of the dynamics?

To see that, we derive the equation of motion with respect to $\{\bm{x}_{e}(\bm{q}),\bm{p}(\bm{q})\}$.
By substituting Eq.~(\ref{eq:xie}) into Eq.~(\ref{eq:EQM}), it
is straightforward to obtain:

\begin{multline}
\left[\begin{array}{cc}
eB_{e}(\bm{q})\hat{z}\times & \frac{\Delta B_{e}(\bm{q})}{B}I\\
-\frac{\Delta B_{e}(\bm{q})}{B}I & -\frac{1}{eB}\left(\frac{\Delta B_{e}(\bm{q})}{B}\right)\hat{z}\times
\end{array}\right]\left[\begin{array}{c}
\dot{\bm{x}}_{e}(\bm{q})\\
\dot{\bm{p}}(\bm{q})
\end{array}\right]\\
=-\mathcal{D}^{\prime\prime}(\bm{q})\left[\begin{array}{c}
\bm{x}_{e}(\bm{q})\\
\bm{p}(\bm{q})
\end{array}\right],\label{eq:EQM2}
\end{multline}
where $\mathcal{D}^{\prime\prime}(\bm{q})$ is the transformed dynamic
matrix with respect to the new dynamic variables.

We observe that the equation of motion becomes more complicated. It
still fits in with the general form of the Sundaram-Niu dynamics,
but with a complicated structure of Berry curvatures~\cite{sundaram1999}.
Similar complexity also arises when one adopts other definitions of
the CP position. It turns out that the initial definition of the CP
position provides the simplest form of equation of motion.

We conclude that alternative definitions of the CP position will not
affect our general interpretation, \emph{i.e.}, CPs follow the Sundaram-Niu
dynamics. The particular choice adopted in Eq.~(\ref{eq:EQM}) is
the best because it has the simplest structure of Berry curvatures.

\section{Numerical simulations}

\subsection{Methods}

We employ the Metropolis Monte-Carlo method to evaluate the coefficients
defined in Eqs.~(\ref{eq:Ak1}, \ref{eq:Dxx}\textendash \ref{eq:Dpp}).
The algorithm and setup of our simulations are similar to those adopted
in Ref.~\cite{yi1998}, with a couple of improvements detailed as
follows.

\begin{figure}
\includegraphics[width=1\columnwidth]{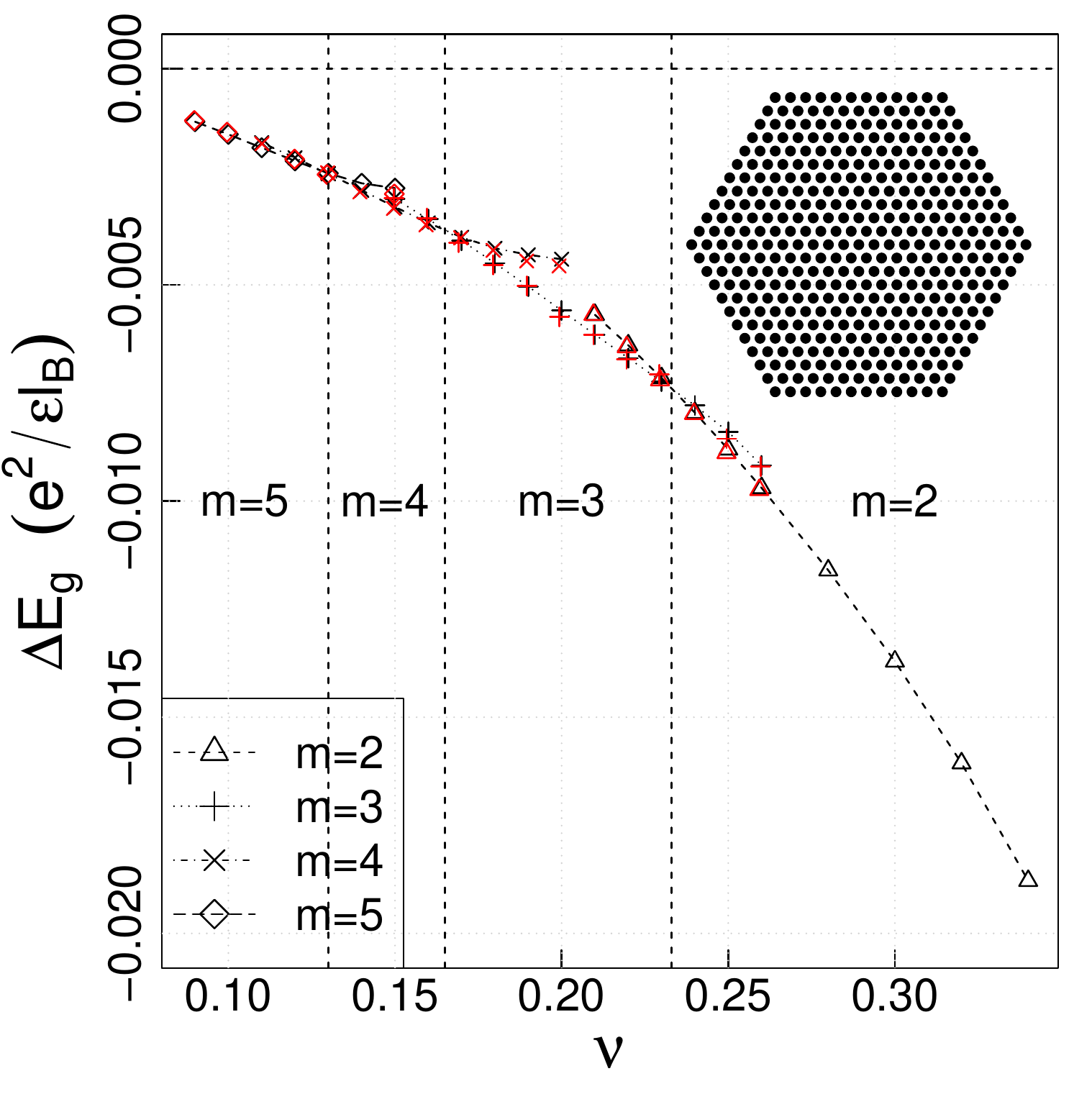}

\caption{\label{fig:Simulation}Variational ground state energies of CPWCs
relative to that of the ordinary WC. Red points indicate the results
of Ref.~\cite{yi1998}. The phase boundaries between CPWC phases
with different values of $m$ are determined by comparing the energies,
and indicated by dashed vertical lines. Inset: configuration of the
simulation cell. }
\end{figure}

Firstly, our calculation employs a much larger simulation cell which
involves 397 electrons arranged as $11$ concentric hexagonal rings
in a plane, as shown in the inset of Fig.~(\ref{fig:Simulation}).
The larger simulation cell is needed to eliminate finite size effects
as the coefficients decay slowly in the real space.

Secondly, we use a different wave function for the finite simulation
cell, and eliminate the need for introducing ``ghost'' particles
explicitly. As pointed out in Ref.~\cite{yi1998}, for a finite lattice
in equilibrium ($\bm{R}_{i}=\bm{R}_{i}^{0}$, and $\bm{k}_{i}=0$),
the average positions of electrons do not coincide with their expected
equilibrium positions due to asymmetry induced by the Bijl-Jastrow
factor. As a result, it is necessary to introduce a cloud of ``ghost
particles'' for each of electrons to counter balance the effect.
In Ref.~\cite{yi1998}, finite size ghost-particle clouds were introduced.
In our simulation, we extend the size of the ghost particle clouds
to infinity. The resulting wave function can be determined analytically:
\begin{multline}
\Psi\left(\left\{ \bm{r}_{i}\right\} \right)\propto\mathcal{A}\frac{\prod_{i<j\leq N}(z_{i}+ik_{i}l_{B}^{2}-z_{j}-ik_{j}l_{B}^{2})^{m}}{\prod_{i\leq N}\prod_{j\ne i,\leq N}\left(z_{i}+ik_{j}l_{B}^{2}-Z_{j}^{0}\right)^{m}}\\
\times\prod_{i=1}^{N}\left[\psi(z_{i}-Z_{i}^{0}+ik_{i}l_{B}^{2})\right]^{m}\phi_{\bm{R}_{i}}(\bm{r}_{i}),\label{eq:wf3}
\end{multline}
where $Z_{i}^{0}\equiv X_{i}^{0}+iY_{i}^{0}$ ($\bm{R}_{i}^{0}\equiv(X_{i}^{0},Y_{i}^{0})$),
$N$ is the total number of electrons in the simulation cell, and~\cite{rhim2015},
\begin{equation}
\psi(z)\equiv\frac{\prod_{i\ne0}\left(z-Z_{i}^{0}\right)}{\prod_{i\ne0}\left(Z_{i}^{0}\right)}\propto\frac{1}{z}\theta_{1}\left(\left.\frac{z}{a}\right|\frac{1}{2}+i\frac{\sqrt{3}}{2}\right),
\end{equation}
where $a$ is the lattice constant of the WC, the product is extended
to an infinite triangular lattice with unit vectors $\bm{a}_{1}=(1,0)a$
and $\bm{a}_{2}=(1/2,\sqrt{3}/2)a$, and $\theta_{1}$ is the Jacobi
theta function. Equation (\ref{eq:wf3}) is used in our numerical
simulations.

An important issue of our simulation is to extrapolate the calculation
results obtained in a finite simulation cell to the macroscopic limit.
To this end, we find that $A_{0}^{\mathrm{finite}}$, coefficient
defined in Eq.~(\ref{eq:A1}) calculated with a harmonic approximation
of the wave function for the finite simulation cell (See Eq.~(\ref{eq:A0})
in Appendix B), fits the long-range tail of the calculated coefficient
very well. Hence, we divide the coefficient into a long range part
$A_{0}^{\mathrm{finite}}$ and a short range part that decays rapidly
with the distance, and fit the short range part up to the fifth nearest
neighbors. The extrapolation is then straightforward by upgrading
$A_{0}^{\mathrm{finite}}$ to its infinite lattice counterpart, which
can be determined analytically. 

Similar extrapolation schemes are applied for the determinations of
the coefficients Eq.~(\ref{eq:Dxx}\textendash \ref{eq:Dpp}). We
can have harmonic approximation for these coefficients as well (see
Eq.~(\ref{eq:D0q}) in Appendix B). They are regarded as the long
range parts of the coefficients. In this case, the remainders of the
coefficients decays as $1/|\bm{R}_{i}^{0}-\bm{R}_{j}^{0}|^{5}$ in
the long range. We fit the remainders of $\mathcal{D}^{\bm{x}\bm{x}}$
and $\mathcal{D}^{\bm{p}\bm{x}}$ with short range terms up to the
fifth nearest neighbors, whereas for $\mathcal{D}^{\bm{p}\bm{p}}$,
the higher precision of the calculated values allows us to fit it
with a $1/|\bm{R}_{i}^{0}-\bm{R}_{j}^{0}|^{5}$ term plus the short
range terms. We note that using the short-range terms to fit the remainders
may yield an incorrect asymptotic behavior in the long wavelength
limit. It makes our determination of the dynamic matrix less reliable
in the regime. 

Figure~\ref{fig:Simulation} shows the variational ground state energies
determined from our simulations, and a comparison with the results
presented in Ref.~\cite{yi1998}. In our simulations, each of Markov
chains contains a total $5.6\times10^{12}$ proposal states with an
acceptance rate $\sim25\%$. They yield essentially identical results
as the old simulations (within the error bars of the old simulations)
albeit with much improved precision.

\subsection{Results}

\begin{figure}
\includegraphics[width=1\columnwidth]{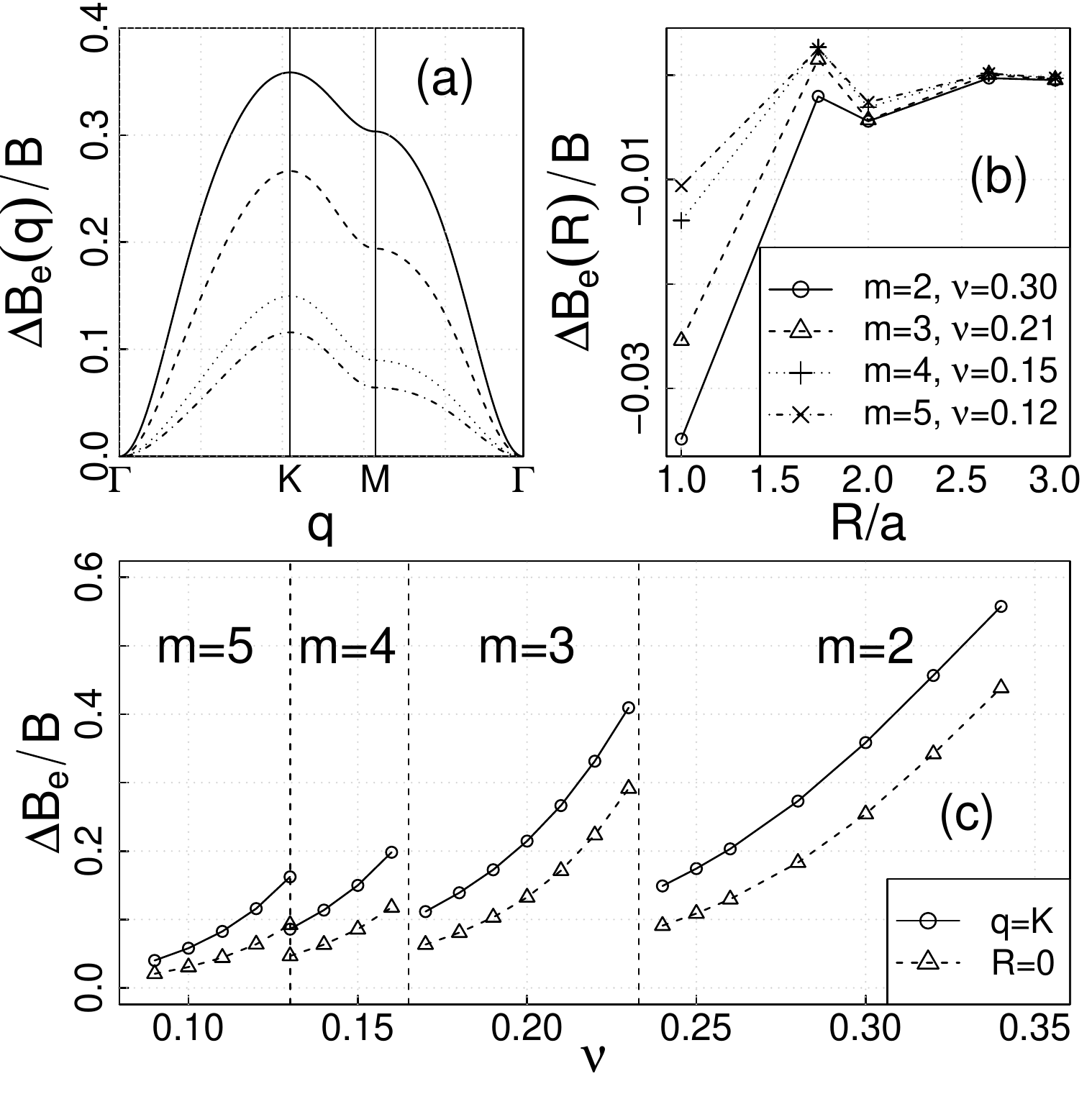}

\caption{\label{fig:Beff}Emergent gauge field. (a) Distribution of the emergent
gauge field in the Brillouin zone for four representative filling
factors (see legends in (b)) with different values of $m$; (b) Decay
of the dissipationless viscosity coefficient $\Delta\mathcal{B}_{e}(\bm{R}_{i}^{0})$
in the real space. $R$ denotes the distance between two particles,
and $a$ is the lattice constant; (c) Filling factor $\nu$ dependence
of the emergent gauge field at the $K$-point of the Brillouin zone
(circle-solid line) and the mean-field value $\Delta B$ defined in
(\ref{eq:DeltaB}) (triangle-dashed line ). }
\end{figure}

Figure \ref{fig:Beff} shows the emergent gauge field $\Delta B_{e}$.
The distribution of the emergent field in the Brillouin zone is shown
in Fig.~\ref{fig:Beff}(a). It peaks at the $K$-point and vanishes
at the $\Gamma$-point. In the real space, the dissipationless viscosity
coefficient decays rapidly with the distance, as shown in Fig.~\ref{fig:Beff}(b).

The strength of the emergent gauge field is characterized either by
the value of $\Delta B_{e}(\bm{q})$ at the $K$-point or the mean-field
value $\Delta B$ defined in Eq.~(\ref{eq:DeltaB}). Both are shown
in Fig.~\ref{fig:Beff}(c). The magnitude of the emergent field is
ranged from a few percents to tens percents of the external magnetic
field, and is an increasing function of the filling factor for a given
value of $m$. The magnitude is smaller than that expected for a FQH
liquid, which has a mean-field value $\Delta B^{FQH}/B=m\nu$. It
indicates the mean-field approximation adopted for the theory of FQH
liquids is not applicable for the CPWCs. On the other hand, the magnitude
is actually gigantic in comparing with that generated by an intrinsic
SOC. For instance, the intrinsic SOC in GaAs could also give rise
to a similar emergent gauge field in an ordinary 2D WC. However, its
magnitude is of order of $\sim0.01\,\mathrm{T}$ only~\cite{ji2017}.

\begin{figure}
\includegraphics[width=1\columnwidth]{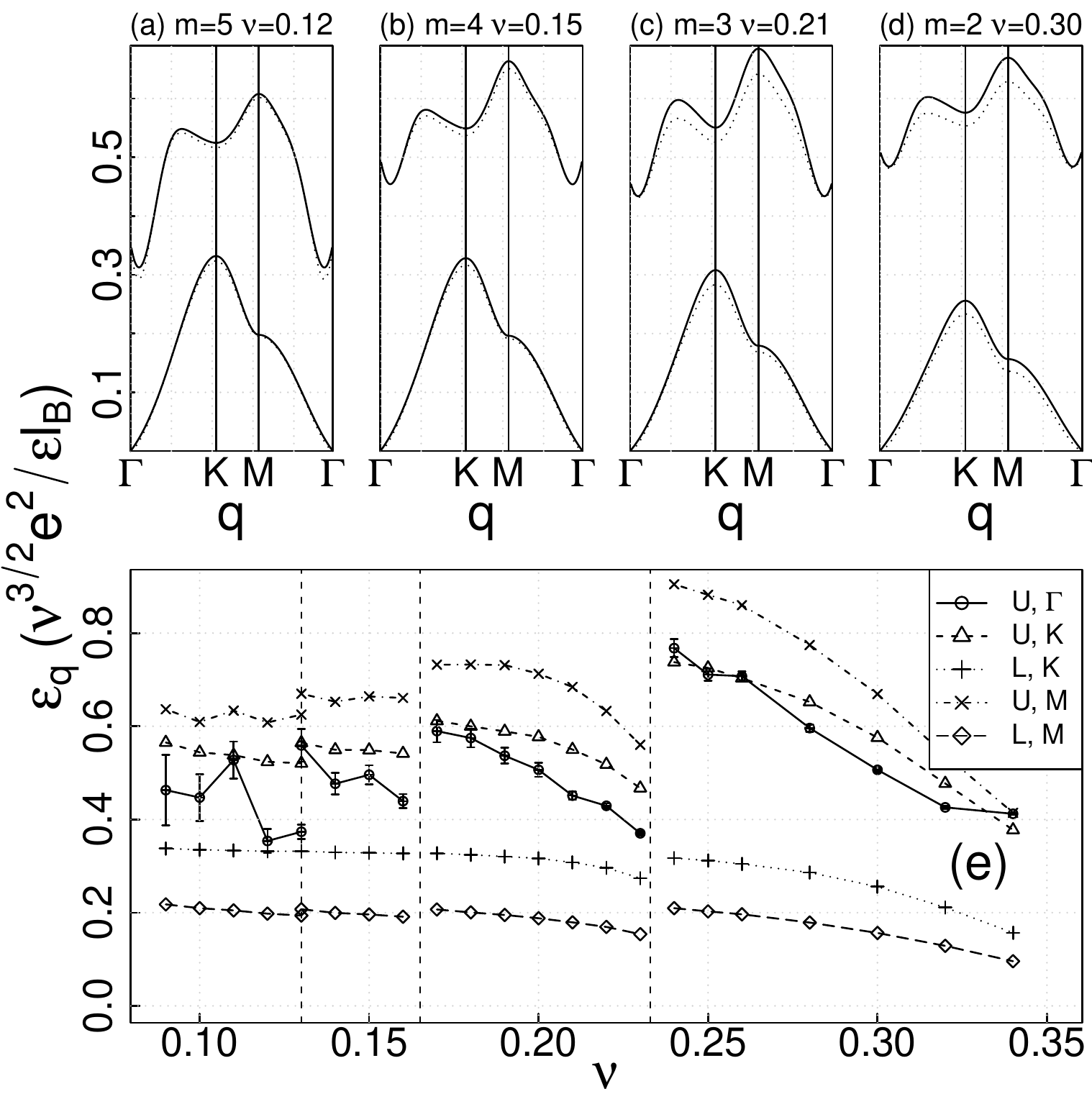}

\caption{\label{fig:Omegaq}Phonon dispersions of type-I CPWCs. (a\textendash d)
Phonon dispersions for a few representative filling factors. Both
results using the projected dynamic matrix (solid lines) and the unprojected
one (dotted lines) are shown. (e) Filling factor $\nu$ dependence
of phonon energies of the upper (U) and the lower (L) branch at high
symmetry points of the Brillouin zone, including $K$ point, $M$
point, as well as $\Gamma$ point (evaluated at $\bm{q}=0.01\bm{K}$).
$e^{2}/\epsilon l$ ($\approx4.3\sqrt{B[\mathrm{T}]}\,\mathrm{meV}$
for GaAs) is the Coulomb energy scale. Error bars for the phonon energies
near the $\Gamma$-point are shown. }
\end{figure}

The phonon dispersions of type-I CPWCs are obtained by solving the
generalized eigenvalue equation Eq.~(\ref{eq:gep}). The results
are summarized in Fig.~\ref{fig:Omegaq}. Among the two branches
of phonons of a CPWC, the lower branch is not much different from
that of an ordinary WC, both qualitatively and quantitatively~\cite{maki1983,c^ote1990,c^ote1991},
whereas the upper branch is an emergent mode with an energy scale
$\sim0.5\nu^{3/2}e^{2}/\epsilon l_{B}$, which is much smaller than
the cyclotron energy. The upper branch has similar origin and energy
scale as the magneto-roton mode arisen in FQH liquids~\cite{girvin1986}.
We thus interpret the mode as the magneto-roton mode of the CPWC. 

\begin{figure}
\includegraphics[width=1\columnwidth]{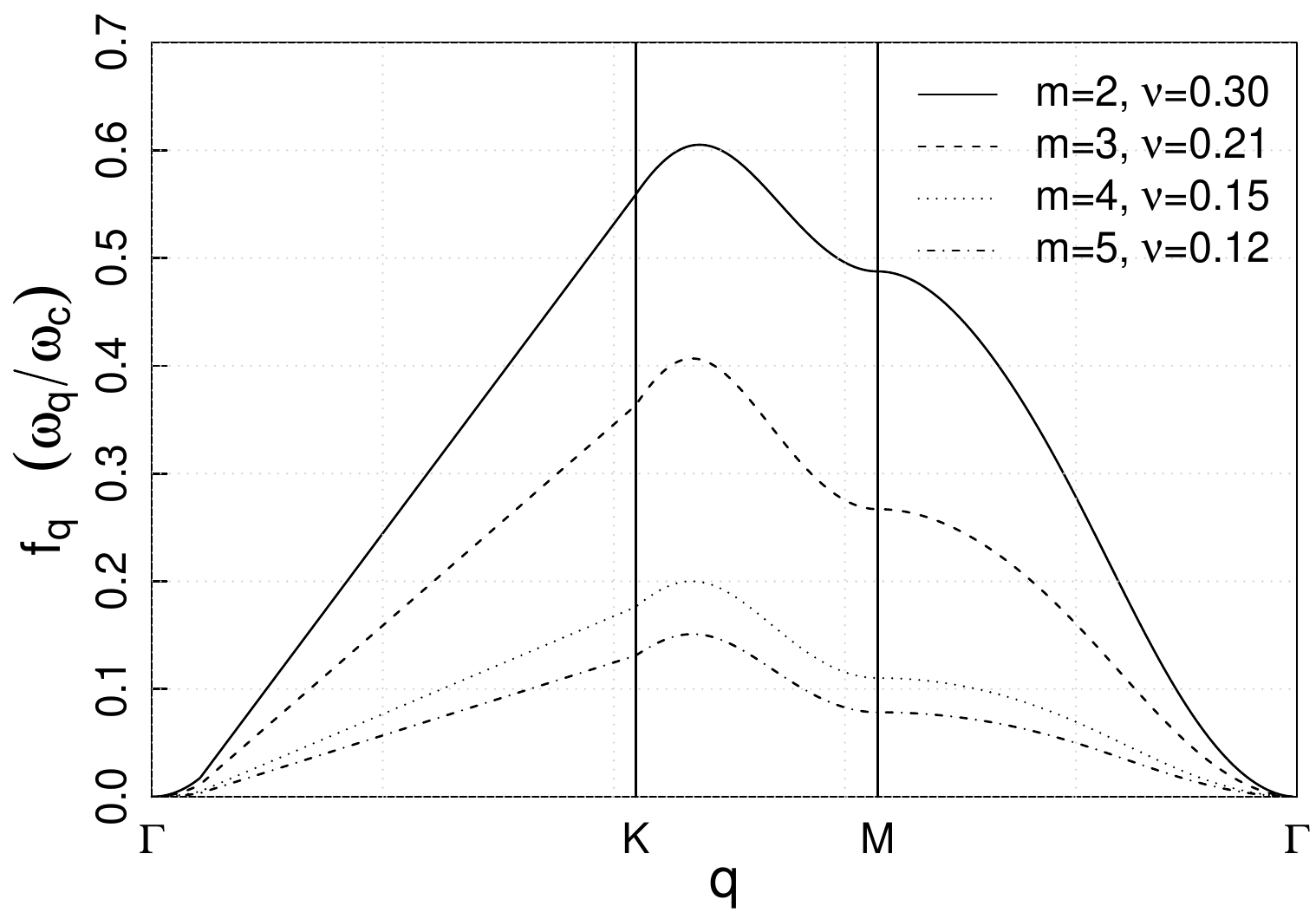}

\caption{\label{fig:Oscillator-strength} Oscillator strength of the emergent
magneto-roton mode for a few representative filling factors, in unit
of $\omega_{\bm{q}}/\omega_{c}$, where $\omega_{\bm{q}}$ is the
frequency of the mode, and $\omega_{c}$ is the cyclotron frequency. }
\end{figure}

We also calculate the oscillator strength of the emergent magneto-roton
mode. To do that, we determine the response of the system to an external
time-dependent electric field $\bm{E}(t)=\bm{E}_{\omega}\exp(-i\omega t)$.
Because the external electric field is only coupled to the electron
degree of freedom (see Sec.~\ref{subsec:Dipole-interpretation}),
it will introduce a scale potential $e\bm{E}(t)\cdot\bm{x}^{e}\equiv e\bm{E}(t)\cdot(\bm{x}-\hat{z}\times\bm{p}/eB)$
into the system. The extra term in the potential could be interpreted
as a coupling to the dipole of a CP. As a result, the equation of
motion has the form:
\begin{multline}
\left[\begin{array}{cc}
eB_{e}(\bm{q})\hat{z}\times & I\\
-I & \frac{1}{eB}\hat{z}\times
\end{array}\right]\left[\begin{array}{c}
\dot{\bm{x}}(\bm{q})\\
\dot{\bm{p}}(\bm{q})
\end{array}\right]=-\mathcal{D}(\bm{q})\left[\begin{array}{c}
\bm{x}(\bm{q})\\
\bm{p}(\bm{q})
\end{array}\right]\\
+\left[\begin{array}{c}
-e\bm{E}(t)\\
\bm{E}(t)\times\hat{z}/B
\end{array}\right].\label{eq:EQM-1}
\end{multline}
By solving the equation, we can determine the displacement of electrons
parallel to the electric field, and the oscillator strength $f_{\bm{q}i}$
is defined by the relation~\cite{johndavidjackson1999} $x_{e}^{\parallel}(\bm{q})=-e/m_{b}\sum_{i}f_{\bm{q}i}/(\omega_{\bm{q}i}^{2}-\omega^{2}-i\omega0^{+})E_{\omega}$,
where $m_{b}$ is the electron band mass of the 2DEG. The oscillator
strength for the emergent magneto-roton mode is shown in Fig.~\ref{fig:Oscillator-strength}.
We see that it vanishes at the limit $\bm{q}\rightarrow0$, consistent
with Kohn's theorem.

The mode we predict here has an energy scale $\sim0.5\nu^{3/2}e^{2}/\epsilon l_{B}$.
For typical experimental parameters, the energy is much larger than
that probed by existing microwave experiments~\cite{andrei1988,engel1997,li1997,li2000,chen2003,chen2004,chen2006,zhu2010,williams1991},
which are focused on the disorder pining modes. Our prediction thus
calls for new microwave experiments to probe the new energy regime. 

\section{Concluding remarks}

In summary, we have derived the effective dynamics of CPs in a CPWC
directly from the microscopic wave function. We find, most notably,
the presence of a Berry curvature in the momentum space. The picture
emerged from the dynamics is different from the conventional CF theory
which assumes that CPs behave just like an ordinary Newtonian particle.
On the other hand, we show that the dynamics is consistent with the
dipole picture of CPs, and the presence of the Berry curvature is
actually an inevitable consequence of the picture. The consistency
is not a coincidence, since both are based on microscopic wave functions.

Although our theory is developed for the CPWC phase, the insight may
be carried over to the liquid phase. In particular, the presence of
the Berry curvature would provide a cure for deficiencies of the conventional
CF theory~\cite{shi2017}. The solution is less radical and would
be more natural compared to that prescribed by the Dirac theory~\cite{son2015}.

Our study reveals the discrepancy between the conventional interpretation
of CPs and that emerged from a microscopic wave function. This is
not surprising because the conventional picture was developed from
a flux-attachment argument for free particles residing in a parabolic
band~\cite{lopez1991}, while the microscopic wave functions are
constructed for electrons constrained in a Landau level. For the latter,
it would be highly desirable to have a CF theory which makes no direct
reference to the magnetic field, since in a Landau level all the effects
of the magnetic field has been accounted for by the Berry curvature.
Indeed, in our dynamics, all the references to the magnetic field
$eB$ could be interpreted as $1/\Omega_{z}$. Such a theory would
also be a first step toward an understanding of the fractional Chern
insulators~\cite{parameswaran2013}.
\begin{acknowledgments}
This work is supported by National Basic Research Program of China
(973 Program) Grant No. 2015CB921101 and National Science Foundation
of China Grant No. 11325416.
\end{acknowledgments}

\appendix

\section{Dynamic matrix}

To derive the formulas for the dynamic matrix coefficients (\ref{eq:Dxx}\textendash \ref{eq:Dpp}),
we make use identities:
\begin{align}
\frac{\partial V_{ee}}{\partial x_{i\alpha}} & =\frac{\partial V_{ee}}{\partial u_{j\beta}}\left[A^{-1}\right]_{j\beta,i\alpha},\\
\frac{\partial V_{ee}}{\partial p_{i\alpha}} & =\frac{1}{\hbar}\left(\frac{\partial}{\partial k_{i\alpha}}-l_{B}^{2}\epsilon_{\alpha\beta}\frac{\partial}{\partial u_{i\beta}}\right)V_{ee},\\
\frac{\partial\left|\Psi\right|^{2}}{\partial u_{i\alpha}} & =\frac{r_{i\alpha}-R_{i\alpha}}{l_{B}^{2}}\left|\Psi\right|^{2},\\
\frac{\partial\left|\Psi\right|^{2}}{\partial p_{i\alpha}} & =\frac{1}{\hbar}\left(\frac{\partial}{\partial k_{i\alpha}}-l_{B}^{2}\epsilon_{\alpha\beta}\frac{\partial}{\partial u_{i\beta}}\right)\left|\Psi\right|^{2}=\frac{l_{B}^{2}}{\hbar}\epsilon_{\alpha\beta}\frac{\partial\left|\Psi\right|^{2}}{\partial r_{i\beta}}.
\end{align}
In deriving the last identity, we make use Eq.~(\ref{eq:DPsi}).
We obtain:
\begin{align}
\frac{\partial V_{ee}}{\partial x_{i\alpha}} & =\frac{1}{l_{B}^{2}}\left\langle \left(\hat{V}_{ee}-\bar{V}_{ee}\right)\left(\hat{r}_{j\beta}-R_{j\beta}\right)\right\rangle \left[A^{-1}\right]_{j\beta,i\alpha},\\
\frac{\partial V_{ee}}{\partial p_{i\alpha}} & =-\frac{l_{B}^{2}}{\hbar}\epsilon_{\alpha\beta}\left\langle \frac{\partial\hat{V}_{ee}}{\partial r_{i\beta}}\right\rangle \,.
\end{align}

Applying the second derivative, and making use the identities again,
we obtain Eqs.~(\ref{eq:Dxx}\textendash \ref{eq:Dpp}).

\section{Harmonic approximations of the coefficients}

A rudimentary approximation for evaluating the coefficients is the
harmonic approximation. We define the harmonic approximation of the
wave function as:
\begin{multline}
\left|\Psi_{0}\right|^{2}\propto\exp\left[2m\sum_{i<j}\ln\left|\bm{r}_{i}-\bm{r}_{j}\right|^{2}-\frac{1}{2l_{B}^{2}}\sum_{i}\left|\bm{r}_{i}-\bm{R}_{i}^{0}\right|^{2}\right]\\
\approx\exp\left[-\frac{1}{2}\sum_{ij,\alpha\beta}F_{\alpha\beta}(\bm{R}_{i}^{0}-\bm{R}_{j}^{0})\xi_{i\alpha}\xi_{j\beta}\right],\label{eq:Psi02}
\end{multline}
where 
\begin{equation}
F_{\alpha\beta}(\bm{R}_{i}^{0})=\begin{cases}
\frac{1}{l_{B}^{2}}\delta_{\alpha\beta} & \bm{R}_{i}^{0}=0\\
-2m\frac{2R_{i\alpha}^{0}R_{i\beta}^{0}-\left|\bm{R}_{i}^{0}\right|^{2}\delta_{\alpha\beta}}{\left|\bm{R}_{i}^{0}\right|^{4}} & \bm{R}_{i}^{0}\ne0
\end{cases}.
\end{equation}

Under the approximation, the coefficient $A$ defined in Eq.~(\ref{eq:A1})
is related to $F$ by a matrix inversion:
\begin{equation}
A\approx A_{0}\equiv\frac{1}{l_{B}^{2}}F^{-1}.\label{eq:A0}
\end{equation}

The coefficient $F_{\alpha\beta}(\bm{R}_{i}^{0})$ satisfies an identity:
\begin{equation}
\sum_{\alpha}F_{\alpha\alpha}(\bm{R}_{i}^{0})=\frac{2}{l_{B}^{2}}\delta_{\bm{R}_{i}^{0},0}.
\end{equation}
It leads to a vanishing emergent gauge field. 

To evaluate the dynamic matrix, we first expand the electron-electron
interaction operator $\hat{V}_{ee}$ to the second order of $\hat{\bm{\xi}}_{i}$.
With the approximated wave function Eq.~(\ref{eq:Psi02}), the coefficients
(\ref{eq:Dxx}\textendash \ref{eq:Dpp}) become Gaussian integrals.
We obtain the harmonic approximation of the dynamic matrix:
\begin{equation}
\mathcal{D}_{0}(\bm{q})=\left[\begin{array}{cc}
D_{0}(\bm{q}) & \frac{l_{B}^{2}}{\hbar}D_{0}(\bm{q})\hat{\epsilon}\\
-\frac{l_{B}^{2}}{\hbar}\hat{\epsilon}D_{0}(\bm{q}) & -\frac{l_{B}^{4}}{\hbar^{2}}\hat{\epsilon}D_{0}(\bm{q})\hat{\epsilon}
\end{array}\right],\label{eq:D0q}
\end{equation}
where $D_{0}(\bm{q})$ is the $2\times2$ classical dynamic matrix
of the WC. 

\nocite{apsrev41Control}

\bibliographystyle{apsrev4-1}
\bibliography{bibcontrol,CFWC_PR}

\end{document}